
\magnification=\magstephalf
\baselineskip=12pt
\nopagenumbers
\parindent=1pc
\parskip=3pt
\font\sixteenrm=cmr17 at 16pt

\def\msun{M_{\odot}}

\def\etal{{\it et~al.\ }}

\def\ie{{\it i.e.,~}}

\def\nl{{\nu_L}}
\def\ltsima{$\; \buildrel < \over \sim \;$}
\def\simlt{\lower.5ex\hbox{\ltsima}}
\def\gtsima{$\; \buildrel > \over \sim \;$}
\def\simgt{\lower.5ex\hbox{\gtsima}}
\def\ref{\noindent\hangindent.5in\hangafter=1}

\vskip 8truepc
\centerline{\sixteenrm  ON THE DISTANCE DETERMINATION AND IONIZATION}
\vskip 1truecm
\centerline{\sixteenrm  OF THE HIGH VELOCITY CLOUDS.}
\vskip 5truepc
\centerline{\bf A. Ferrara$^{1,3,4}$ {\rm and} G. B. Field$^{2}$}
\centerline{\it ${^1}$ Space Telescope Science Institute, 3700
San Martin Drive, Baltimore, 21218 MD}
\centerline{\it ${^2}$ Harvard-Smithsonian Center for Astrophysics, 60
Garden Street, Cambridge, 02138 MS}
\centerline{\it ${^3}$ Osservatorio Astrofisico di Arcetri,
Largo E. Fermi, 5, 50125 Florence, Italy}
\centerline{\it ${^4}$ Affiliated with the Space Science Department, ESA}
\vfill\eject
\centerline{\bf ABSTRACT}
\vskip 1truecm
We present a study of the ionization and thermal structure of  neutral
hydrogen clouds located in the Galactic halo, immersed in the extragalactic
background radiation field, and supposed to be in pressure equilibrium with
the surrounding (presumably) hot medium. The problem is solved numerically,
but an useful analytical approximation has been derived as well. We discuss
the two main parameters of the problem, \ie the radiation field and the halo
pressure, in the framework of the current models for the halo/disk interaction.
We find a well defined relation between a critical column density at
which the cloud starts to develop a cold central core with the cloud linear
size.  Making use of this relation, we suggest a straightforward method to
derive the distance to the cloud. We discuss the possible sources of error
in this determination and it is found that the method is particularly suitable
for those clouds which are subsonic with respect to the surrounding medium,
while for the supersonic ones, the method can only give a lower limit to
the distance. The H$\alpha$ emission from the partially ionized edge of the
cloud is calculated and compared with the available observations; this
measure is a powerful indicator of the possible presence of a shock in the
cloud.
\vfill\eject
\centerline{\sl 1. INTRODUCTION}
\medskip
One of the most long standing problems in the study of the
interstellar medium is represented by the so-called high velocity clouds
(HVCs). Generally speaking, HVCs were  hystorically defined as clouds
of neutral gas with a velocity relative to the LSR $\vert v_{LSR}
\vert\simgt 90$~km~s$^{-1}$. Here we use the term in a
somewhat looser manner, since we mean the HI cloud with velocities
incompatible with a simple model of galactic rotation and structure.
The only requirement is that the clouds are located well above the
main Galactic disk into the halo.
The mass flux on the galactic disk provided by the HVCs and the
VHVC, very high velocity clouds, is about $0.5 \msun $~yr$^{-1}$
(Mirabel 1989); Sembach \etal (1991) and Wakker (1991) derive a value
of about $5 \msun $~yr$^{-1}$ including all HI with velocities not fitting
galactic rotation. An estimate of their metallicity given by
de Boer \& Savage (1984) is $Z\le 0.3 Z_\odot$; similar results are
obtained by Blades \etal (1988) for two HVCs in the direction of
SN1987A. HVCs are usually observed in HI 21~cm line emission; however
in the last years a number of different observations have been
attempted. Colgan \etal (1990) (and references therein) have reported
a negative result in detecting the 21~cm absorption line, which  they interpret
as a requirement for an external heating source. Wakker \& Boulanger
(1986) have used the IRAS data to search for $100 \mu$m emission from
dust inside the HVCs but none of the sample clouds has been detected.
This implies a lower-than-normal dust content or a low temperature of
the dust. There are no direct evidences of molecular hydrogen related
to the HVC, even if Rohlfs \etal (1989) suggest that the high latitude
molecular cloud G90+38 can be spatially connected with an infalling HVC.
Absorption lines towards halo stars and extragalactic objects may provide
useful information on the composition of the HVCs. However the only
firm detection of a HVC in a stellar spectrum is towards the star
HD 135485 (Albert \etal 1989). Very recently, Bowen \& Blades (1993)
have detected for the first time Mg~II absorption lines from two
HVCs, which are part of the Complex C, in the line of sight of Mrk~205.
Their results suggest a very different column density ratio N(Mg~II)/N(HI)
in the two clouds, which might be explained by a real difference in the metal
content of the two clouds or by a difference in the ionization
conditions: it is important to note at this regard, that the HI column
densities for the two clouds differ roughly by a factor $~10$
(typical HI column densities for HVCs are in the range $ 10^{18} <
N_H < 10^{21}$~cm$^{-2}$). A relevant feature, discovered through the high
resolution synthesis maps (Wakker \& Schwarz 1991), is the presence
of bright, dense cores with $N_H$ well in excess of $10^{20}$~cm$^{-2}$.
A type of structure for the clouds made by bright concentrations
with low velocity widths ($\sim 7$~km~s$^{-1}$) and extended envelopes
with  much larger widths ($\sim 23$~km~s$^{-1}$) was already indicated by
Cram \& Giovanelli (1976).
\par
As can be realized from the previous discussion, our understanding of the HVC
phenomenon is quite fragmentary and incomplete. Definitely, the major
source of ignorance is the lack of a precise distance estimate.
The central importance of the distance
knowledge has been stressed by many authors (Van Woerden 1993 and references
therein).
{}From the observational point of view, the claim made by Danly, Albert \&
Kuntz (1993) of the bracketing of the distance to a HVC in the Complex M
in the range $1.5<z<4.4$~kpc, provides
at least the first firm evidence that the HVCs are located in the Galactic
halo. The distance estimate is particularly important because most
of the physical parameters  of the clouds scale with the distance $\Delta$:
the physical size, $\ell\propto \Delta$, the density, $n\propto \Delta^{-1}$
and the mass, $M\propto \Delta^2$. In addition the related problem of the
origin of the HVCs could also strongly benefit  from such an information.
The most common explanation for the origin of the HVCs is the so-called
galactic fountain model (GF) worked out by Shapiro \& Field (1976) and
subsequently detailed by Bregman (1980); Houck \& Bregman (1990) extended
it to the low regions of the halo; Li \& Ikeuchi (1992) have investigated
their formation in giant halos. In this model the HVCs correspond to the
condensation mode of a thermal instability occurring in a flow of hot gas,
generated by supernova explosions, rising from the disk. Altough the GF
can explain most of the data concerning the HVCs (but not all: see the
discussion
in Wakker 1989), many doubts are still present on the reality of the
entire circulation process. In fact, it is not completely clear
that the superbubbles are able to breakout of the disk when the thick
(``Lockman'') exponential layer of the HI disk is considered
(Mac Low \& Mc Cray 1988; Mac Low \etal 1989; Tenorio-Tagle \etal 1990),
and particularly when the effects of a magnetic field inhibiting the growth
of the supernova shock-sweeped region are taken into account (Tomisaka 1990;
Ferriere \etal 1991; Shapiro \& Benjiamin 1993; Norman 1993).
On a different basis Ferrara \&
Einaudi (1992) pointed out that, under the regime prevailing in a fountain
flow, dynamical instabilities, leading to convective motions rather than
non-gravitational condensations, may have a faster growth rate, thus
quenching the cloud formation process.  On the other hand, clear evidence
of hot gas ($T_h\simgt 10^6$~K) located in the Galactic halo comes from
the ROSAT shadowing experiments towards the Draco cloud (Burrows \& Mendenhall
1991; Snowden \etal 1991; Herbstmeier \etal 1993). Also, Herbmeister \etal 1993
report an enhancement of the soft X-ray emission near some HVCs in the northern
sky. This result brings fresh support to the existence of an
extended hot halo and poses some constraints on its physical characteristics.
It is probably redundant to stress at this point the value of the distance
determination in order to make some progress on the origin of the HVCs: if they
are located in the halo they must be in pressure equilibrium with the
hot gas in order not to be rapidly dispersed; furthermore the study of this
interaction may
lead to a better comprehension of many different phenomena (accretion
of material onto the galactic disk, halo and disk star formation, Galactic
chemical evolution, general gas circulation, structure of the halo)
that are still unclear.
\par
Finally, if HVCs are located well above the main gaseous Galactic disk, they
are exposed to the ionizing extragalactic background radiation field (EBR).
For the purpose of obtaining a determination of this field at redshift $z=0$,
the HVCs are an  ideal ``test particle'', being a relatively quiet and little
contaminated environment. This point have already been
stated by other authors (Cowie \& McKee 1976; Cowie \& Songaila 1986;
Songaila, Cowie \& Weaver 1988); in addition,
the local value of the EBR has been recognized to  be crucial in
constraining many cosmological models. The principal response of the clouds
to the ionizing radiation is the $H\alpha$ recombination emission. A number
of searches of this emission have been performed in the last years:
at least one positive detection has been reported (Kutyrev \& Reynolds 1989).
\par
In this paper we are proposing a new method to derive the distance to the
HVCs based on a photoionization model for the clouds, supposed to be irradiated
by the EBR.
We will show that the observed core/envelope structure of many HVCs can
be used to derive the cloud linear size, and, from the knowledge of
the angular size, the distance to the cloud can, in principle, be obtained.
Of course, there are posible sources of error in the method either of
theoretical and observational nature, and we discuss the first ones and
point out the possible observational difficulties.
Also, matching the $H\alpha$ emission
deduced from the model to the observed one, further insight on the physical
state of the cloud can be obtained.
\par
The structure of the paper is as follows. In \S~2 we describe the
adopted photoionization model  and discuss its relevant parameters;
\S~3 is dedicated to the results and to the distance estimation method,
while in \S~4 some additional implications are discussed.
\medskip

\bigskip
\centerline{\sl 2. IONIZATION MODEL}
The equations governing the
steady-state ionization and thermal balance of a cloud immersed in an
isotropic radiation field of intensity $I_{\nu}$ are:
$$n(X^i)\int_{\nu_{L_X}}^{\infty} {J_{\nu}\over h \nu}
\sigma_\nu(X^i) d\nu \left[1+\phi(X^i)\right]+ \gamma_c(X^i,T)
n(X^i) n(e) = n(X^{i+1}) n(e) \alpha (X^i,T),\eqno(2.1)$$

$$ \sum_i \sum_X n(X^i)\int_{\nu_{L_X}}^{\infty} {J_{\nu}\over h \nu}
h[\nu-\nu_{L_X}]\sigma_\nu(X^i) d\nu = \sum_i \sum_X {\cal L}
(X^i),\eqno(2.2)$$

$$ n(e) = \sum_{i} \sum_{X} i n(X^i), ~~~~~~~~~~~i=0,......\eqno(2.3)$$

$$P=kT(n(e)+\sum_i \sum_X n(X^i)).\eqno(2.4)$$
These equations express the ionization and thermal balance, and charge
conservation, while eq.(4) is the equation of state for the gas;
$X$ denotes the element considered and $i$ its state of ionization;
$J_\nu$ is the first moment of the field, $\nu_{L_X}$ indicates the
ionization limit for each species,
$\sigma$ is the photoionization cross-section,  $\phi$ is the
secondary ionization rate, $\gamma_c$ is the collisional ionization
coefficient, ${\cal L}$ indicates the appropriate volume cooling rate,
and $\alpha$ is the total recombination coefficient.
We have adopted the ``on the spot'' approximation in which
the diffuse field photons are supposed to be absorbed close to the point
where they have been generated.
\par
The various elements are divided in ``primary'', which enter the
ionization-thermal equilibrium equations (2.1)-(2.4)  and ``secondary''
which do not.
In the following, H and He are considered as primary, while other
elements (C,N,O,Si,Fe)
just contribute to the cooling function and to the electron density (C only).
Secondary elements are considered to be completely ionized by the UV field
below 13.6 eV; a metallicity $Z=0.25 Z_{\odot}$ and a helium abundance
equal to 0.1H as been assumed throughout the paper. Double ionization of
He has been neglected, and He fractional ionization has been supposed equal
to the H one, $x$; the ionization cross section for He has been taken from
Brown (1971). Helium is not of special importance in the temperature range
that we consider; this justifies the rough approximation adopted.
Secondary ionization rates and fractional heating
for H and He have been taken from Shull \& Van Steenberg (1985).
The following processes have been included in the calculation of the
cooling function: i) free-free from all ions; ii) H and He recombination;
iii) electron impact ionization of H and He; iv) electron impact excitation
of H and He (n=2,3,4 triplets); v) He dielectronic recombination; vi)
electron and H impact excitation of secondary elements; excitation
of the metastable levels by electron impact are also included.
The obtained cooling function is almost identical to the one given
by Dalgarno \& McCray (1972).
In additon to photoionization, heating is also provided
by C ionization (2 eV per ionization), assuming that the recombination
coefficient $\alpha$  for C is equal to the H one (Spitzer 1978).
The numerical values for the various coefficients have been taken from
Black (1981) and  Dalgarno \& McCray (1972).
\par
In order to obtain the spatial ionization and thermal structure of the cloud,
eqs. (2.1)-(2.4) must be solved simultaneously with the radiation transfer
equation. We have postulated that the cloud can be modeled as a slab of gas
of thickness $\ell$ illuminated on both sides by the (isotropic) radiation
field  $I_\nu$. With the above assumption the transport equation reads
$$ {dI_\nu(z) \over dz} = -(n(H)\sigma_\nu(H)+n(He)\sigma_\nu(He))I_\nu(z)
\eqno(2.5)$$
which gives the solution
$$J_\nu(z)=2 \pi I_{\nu,0}\left[ E_2(t_\nu^+(z)) +
E_2(t_\nu^-(z))\right]\eqno(2.6)$$
where $E_2$ is the exponential integral function, $I_{\nu,0}$ is the field
intensity at the cloud edge, and
$$ t_\nu^+(z)=\int_{z}^{\ell/2} (n(H)\sigma_\nu(H)+n(He)\sigma_\nu(He))(1-x)
dz\eqno(2.7)$$
$$ t_\nu^-(z)=\int_{-\ell/2}^{z} (n(H)\sigma_\nu(H)+n(He)\sigma_\nu(He))(1-x)
dz\eqno(2.8)$$
\par
The system of equations (2.1)-(2.5) can be numerically solved trough an
iterative scheme. We have requested the pressure $P$ to be constant
throughout the cloud; this condition is particularly suitable to describe
the HVCs, which are believed to be in pressure equilibrium with a hot
external medium. This argument, initially introduced by Spitzer (1956),
is based on the fact that the pressure of the disk intercloud gas at 1 kpc
would be far too low to confine the cloud, which would rapidly expand
decreasing the density to values not allowing the detection of any
spectral absorption feature toward distant stars.
\par
In analogy with the usual HII regions,
we expect to find a critical length, $\ell_c$, analogous to the
usual Str\"omgren radius. The latter can be dimensionally
evaluated equating the number of available ionizing photons to the
number of recombinations inside a sphere, but in principle one can
calculate the structure of the partially ionized zone at the interface
between the cloud and the surrounding medium using some analytical
approximations for eqs. (2.1)-(2.5). If we neglect secondary and collisional
ionizations, the electron contribution due to secondary elements and helium,
one can write the ionization equation in the isobaric case and for a power-law
exciting spectrum of the form $J_\nu= J_0(\nu/\nu_L)^{-\gamma}$ as
$${x^2\over (1-x)}= {J_0\int_\nl^\infty d\nu {\sigma_\nu\over h\nu}
({\nl\over\nu})^\gamma\over  n \alpha}={J_0\zeta (1+x)kT\over
P \alpha},\eqno(2.9)$$
where $\zeta$ denotes the integral.
In terms of the ionization parameter $\Xi=P/J_0\zeta$, eq. (2.9) becomes
$${x^2\over (1-x^2)}= {kT\over \Xi \alpha};\eqno(2.10)$$
the transport equation (2.5) can also be simplified to give
$${dJ_0\over dz}=-\left[\int_\nl^\infty d\nu \sigma_\nu (1-x)n\right]J_0
\simeq -{1\over 4}\sigma_\nl {(1-x)\over (1+x)}{P \over k T}J_0,\eqno(2.11)$$
where $\sigma_\nl=\sigma_\nu(\nu_{L_H})$.
Substituting for $J_0$ and ${dJ_0/ dz}$ from eq. (2.9)-(2.11) we obtain
the equation for the spatial behavior of the ionization fraction:
$$\int_{x_0}^x {dx \over (1-x)^2x}=-\int_0^z dz{\sigma_\nl P\over 8kT},
\eqno(2.12)$$
where $x_0=x(z=0)$ can be directly derived from eq.(2.10)
$$x_0=\sqrt{ 1\over  1+(\Xi \alpha/kT)}.\eqno(2.13)$$
Since both the integrand and the integral limits of eq. (2.12) depend on
the temperature, some additional approximation is required. To estimate
the value of the temperature we run the numerical code for an optically thin
case to calculate
the (constant) value of the temperature $T_0$ to be substituted in
eq. (2.13) as a function of $\Xi$. In addition, in order to integrate
eq. (2.12), we assume that the temperature inside the cloud remains
approximately constant at the value $T_0$. Since most of the cooling
mechanisms depend on the electron density $n_e$, this simplification is
fairly good for a relatively high ($x\simgt 0.70$) ionization fraction.
In this case we find that the equation of state of the gas closely resembles
a polytropic law
$$P\propto \rho^{(n/n+1)},~~~~~~~~n=-6.25,\eqno(2.14);$$
the properties of polytropes with negative indexes have been studied
by Viala \& Horedt (1974).
The solution $x(z)$ of eq. (2.13) is then  implicitly given by
$$\log{x\over |x-1|} - {1\over(x-1)}= C_1 - {z\over \lambda },\eqno(2.15)$$
where $C_1=\log{x_0/ |x_0-1|} - (x_0-1)^{-1}$ and now $x_0$ is calculated
for $T=T_0$ obtained numerically. Fig. 1 shows various solutions of eq.(2.15)
for different values of the ionization parameter, for a spectral index
$\gamma=1.4$.
\par
The main feature of the solutions is that, as
one may expect, $\ell_c$ tends to become smaller for large values
of  $\Xi$ but the front is smoother.
Also, for low values of $\Xi$,  the ionized zone can be as deep as
several Kpc, and that means that in {\it low pressure regions HVCs are not
expected to have a dense, cold core}. Although only approximate,
the solutions found can be used to gain a qualitative insight of the
problem and may serve as a test for the numerical scheme.
\par
In the following we will discuss the two main input parameters of
the problem, namely the radiation field and the pressure of the
Galactic halo.
\par
\centerline{\it 2.1 Halo radiation field}
\par
The ionizing radiation field in the halo is the sum  of different
components: the stellar radiation field from the disk, the quasar
component and the X-ray background. As far as the disk contribution
is concerned we will neglect it for the following reason. According
to Brown \& Gould (1970) (there is no appreciable difference at low
energies with respect to the improved photoelectric cross-sections given
by Morrison \& Mc Cammon 1983) the energy at which half of the radiation
is absorbed by gas in the Galactic disk is
$$\left(E^*\over {\rm keV}\right)=\left(N_H\over 5.9 \times 10^{21}{\rm
cm}^{-2}\right)^{3/8}.\eqno(2.16)$$
If we consider the value of $N_H=3.1\times 10^{20}$~cm$^{-2}$ appropriate for
the thick HI exponential distribution as given by Dickey \& Lockman (1990),
then $E^*=0.33$~keV. Thus, unless the HI is very clumpy (a claim that  has
no actual observational support, Jahoda \etal 1985), it seems very unlikely
that any source of hard photons located in the stellar disk may contribute
sensibly to the halo radiation field. Photons from a population of hot stars
like the central stars of planetary nebulae and white dwarfs with a larger
scale height could be able to penetrate the layer, but their number is quite
uncertain. In addition, even if present in appreciable number, these photons
are barely sufficient to maintain the ionization of the electron component
that is known to extend out to about $\sim 1.0$~kpc as pointed out by
Reynolds (1990),(1993) and Nordgren, Cordes \& Terzian (1993). In this
sense the electron component almost totally insulate the halo from the
Galactic ionizing radiation. The same conclusion can be drawn if one considers
photons escaping from the open chimneys (Norman \& Ikeuchi, 1989): again
the majority of the photons are needed to maintain the ionization of the
electron layer (Norman \& Panagia 1991). Thus it appears that works including
the contribution from photons escaping in the halo from the interior of
multi-supernova remnants (Bregman \& Harrington 1986) are overestimating
the ionizing flux.
\par
A model for the radiation field in the halo proposed by Fransson \& Chevalier
(1985) has been widely used. The spectrum given by those authors, is
the sum of the quasar component obtained by Sargent \etal (1979) using the
Schmidt
luminosity distribution and the extragalactic X-ray background measured
by Schwartz (1979).
Lately, new studies have become available either theoretical and observational.
While the X-ray portion of the spectrum given by Schwartz (1979) has been
confirmed by recent observations by {\it Ginga} (for a review see Mc Cammon
\& Sanders 1990), for what concerns the QSO radiation field
the newly obtained quasar luminosity functions have produced different
estimates of the QSO contribution to the local UV extragalactic background
(Terasawa 1992, Madau 1992). In particular,  Madau (1992) includes the
effects of the opacity associated with the intervening Ly$\alpha$ clouds
and and Ly-limit absorption systems. Therefore, when these results are
considered for the UV region of the EBR, we obtain the adopted spectrum
$$I_\nu(E)=\cases{ 5.96\times 10^{-24} E_{Ry}^{-1.4}
&$13.6 eV\le E \le 1.5 keV$;\cr\cr 5.1\times 10^{-26}E_{keV}^{-0.4}e^
{-E/41~keV}&$1.5 keV\le E$.~~~~~~~~~~~~~~~~~~~~~~~~~~~~~~~(2.17)\cr}$$
Note that this value is in agreement with the estimate
$I_{\nu}(Ry)\simeq 6\times 10^{-24}$ derived by Kulkarni \& Fall (1993)
from a proximity effect study in the distribution of Ly$\alpha$ forest
lines at low redshift.
\medskip
To take into account the possible radiative contribution
from the hot confining gas, we have also used a composite spectrum
which is a sum of the EBR and the hot gas emission contributions.
For a plasma temperature $T_h\sim 10^6$~K the continuum emission is
dominated by the free-free processes (Landini \& Monsignori Fossi 1990). The
intensity of the free-free emission can be written as
$$I_\nu^{ff}(E)= 1.66\times 10^{-20} (EM) T_h^{-0.5} g(\nu, T_h) \sum {Z^2 n_z
\over n_H}e^{-E/kT_h}, \eqno(2.18)$$
where $EM$ is the emission measure, $T_h$ the temperature of the
plasma, $g(\nu, T_h)$ is the Gaunt factor, $Z$ is the ion charge and
$n_z/n_H$ its abundance; following Landini \& Monsignori Fossi (1990),
we have included only H and He.  The adopted values for the hot gas are
taken from Burrows \& Mendenhall (1991) who have derived $T_h=1.25\times
10^6$~K and $EM=0.006$~cm$^{-6}$~pc.
\par
However, from an inspection of Fig. 9 of Landini \& Monsignori Fossi (1990),
it appears that, at the $T_h$ of interest, the total cooling losses are
dominated by line emission, whose emissivity is about 18 times larger than
the continuum one. In order to take into account this aspect, although in
a rough manner, we assume that the spectral {\it shape} of the total emission
from the hot gas (continuum+lines) is the same as the free-free but with
a coefficient larger by a factor 18. This should be a reasonable assumption
since the overwhelming majority of the lines for such a plasma are in the
region shortwards of $912$~\AA.
\par
\centerline{\it 2.2 Halo pressure}
\par
The determination of the pressure the Galactic halo is a difficult
task    because it involves a firm  understanding not only of the
disk/halo interaction, but also of the processes occurring at the
interface with the intergalactic medium.
A great deal of uncertainty, as already discussed in the
Introduction, persist on the nature (if any) of the disk/halo gas
circulation. This topic has been recently reviewed by Spitzer (1990),
who has also pointed out that the relationship among quiet phenomena
as the buoyant gas in a fountain and highly dynamic transient ones
is far to be clear. In addition, the question if the halo can be
understood on average as a steady, and perhaps even static, structure
still remains unanswered; nevertheless, many authors have investigated
one-dimensional hydrostatic models for the halo (Fransson \& Chevalier
1985; Hartquist \& Morfill 1986; Bloemen 1987; Boulares \& Cox 1990).
The purpose of this \S~ is to use some of these results together with
simple estimations to obtain reasonable limits to the halo pressure.
\par
\centerline{\it 2.2.1 Dynamical models}
\par
Recently, Li \& Ikeuchi (1992), have investigated the formation of
giant halos around spiral galaxies. Their paper provides useful information
about the pressure distribution in dynamically, fountain-dominated halos.
Among the three types of halos they find (wind, bounded and cooled)
only the last one (cooled)  forms HVCs by thermal instabilites in a
region between $0\le \varpi \le 20$~kpc and $0\le z\le 10$~kpc,
well out of the main galactic disk.
The density and temperature vary quite rapidly in this region
as can be realized from their Figs. 8(a)-8(b) and, therefore, rather than
looking for a regular pattern we just derive an upper and lower limit to the
pressure:
$$1.3\times 10^{-17} \le P \le 2.3\times 10^{-14}{\rm
ergs~cm}^{-3};\eqno(2.19)$$
all the intermediate values are present in the cloud forming layer.
This halo corresponds to a a temperature and density at the disk
$T_0=10^6$~K and $n_0=5\times 10^{-3}$~cm$^{-3}$, therefore $\log P_0/k=
3.7$
It has to be
pointed out that the pressure derived by Li \& Ikeuchi  is
generally much lower than the one of the intergalactic medium (see next
section), whose presence they did not consider.  This could represent an
inconsistency for their model.
\par
\centerline{\it 2.2.2 Static models}
\par
The static model can be ideally divided in ``hot'' and ``cold'' ones,
\ie in which the gas support comes from its own thermal pressure or form
pressure contained in other phases (magnetic fields, cosmic rays, turbulent
motions), respectively.
\par
The simplest model for a hot halo is the isothermal one.
The 2D pressure distribution in such a model
can be derived from the hydrodynamic in a straightforward manner.
The static Bernoulli equation can be written as
$$c_s^2 \log \rho(\varpi,z)+\phi(\varpi,z)=\psi(\varpi),\eqno(2.20)$$
where $\phi$ is the Galactic gravitational potential, $\varpi$ is the
galactocentric radius and $\psi$ is an integration constant with respect to
$z$ that can be obtained from a comparison with the $\varpi$--component of
the steady momentum equation.
For a constant galactic circular velocity $v_\varphi$,
$\psi(\varpi)=v^2_\varphi \log(\varpi) + C_2,$
where $C_2$ is an integration constant. Therefore
$$P(\varpi,z)=P_\odot \left({\varpi \over  \varpi_\odot}\right)
^{\beta} e^{(\Phi_\odot -\Phi(\varpi,z))/c_s^2},\eqno(2.21)$$
where $\beta={v_\varphi/c_s}^2$.
In order to avoid an unphysical result for $\varpi \rightarrow
\infty$ we must assume that far away $\beta\to 0$. This leads
to the final expression
$$P(\varpi,z)=P_\odot e^{(\Phi_\odot -\Phi(\varpi,z))/c_s^2}.\eqno(2.22)$$
\par
When the limit
$$\lim_{z\to \infty} P(\varpi,z)= P_\odot e^{\Phi_\odot/c_s^2}=P_{IG}$$
where $P_{IG}$ is the intergalactic pressure, is taken, given the values of
$P_{IG}$ and $P_\odot$, $c_s$ is uniquely determined.
The value of the gravitational potential at the solar radius is $\Phi_\odot=4.8
\times 10^{14}$~cm$^2$~s$^{-2}$ for a circular velocity
$v_{\varphi}(\varpi_\odot)= 220$~km~s$^{-1}$ (Binney \& Tremaine 1987).
Assuming for the intergalactic medium $T_{IG}=10^8$~K and $n_{IG}=10^{-6}$~
cm$^{-3}$, and for $P_\odot=3.5\times 10^{-12}$~ergs~cm$^{-3}$ (Cox 1990),
the derived temperature is $T\sim 10^6$~K, exactly
the same used as a boundary condition  by the dynamical model of Li \&
Ikeuchi (1992).
A plot of the pressure isocontour obtained from eq.(2.22) is shown in Fig. 2.
The assumed potential is taken from de Boer (1990), also used by Ferrara
(1993) to explain the vertical equilibrium of the Lockman component.
It consists of a Oort vertical distribution and an exponential stellar
radial potential:
$$\Phi(\varpi,z)=\sigma^2_g(\varpi)\left[\log  \cosh(z/z_0) +
{1\over 2}\epsilon(\varpi)(z/z_0)^2 \right]\eqno(2.23)$$
where $z_0=250$~pc, $\sigma_g^2(\varpi)=(15.4$~km~s$^{-1})^2 \exp {(\varpi_
{\odot}-\varpi)/0.44\varpi_{\odot}}$, and $\epsilon=0.04, 0.07, 0.14$ for
$\varpi=5, 10, 15$~kpc, respectively. A quadratic interpolation to
$\epsilon$ has been used whenever required.
In the region of interest ($0\le \varpi \le 20$~kpc, $0\le z\le 10$~kpc)
the lower and upper limits for a hot static model from Fig. 2 are
$$ 10^{-15} \le P \le  10^{-12} {\rm ergs~cm}^{-3}\eqno(2.24).$$
\medskip
In the cold halo of models the support in the gravitational field
is provided by sources different from thermal energy, as discussed above.
We will refer, as a prototype, to the one-dimensional model
presented by Boulares \& Cox (1990), which appears to include the most updated
data compilation of the  distribution of the various phases of the ISM.
A fit of the exponential component (which is the important one at high $z$)
of the pressure from their results gives the following expression for the
solar neighborhood
$$P(z)=2.61 \times 10^{-12} e^{-z/1.23 {\rm kpc}}{\rm ergs~cm}^{-3}.
\eqno(2.25)$$
We use this result in the same region considered for the Li \& Ikeuchi
model ($3\le z \le 10$~kpc). In this region the lower and upper limits
for the Boulares \& Cox cold static model from eq.(2.25) are
$$ 2.2\times 10^{-13}\le P\le 7.7\times 10^{-16}{\rm
ergs~cm}^{-3}.\eqno(2.26)$$
\par
The range of pressures obtained for the three different (dynamical, static
hot/cold) types of halos is large and hence pretty unsatisfactory.
This fact reflects the extremely poor comprehension of the halo and strongly
recommends futher theoretical and observational studies.

\bigskip
\centerline{\sl 3.  RESULTS AND THE DISTANCE METHOD}
\smallskip
The method we are proposing in order to determine the distance to the
HVCs is essentially based on their detailed ionization structure, which is
completely determined by  the 5 coupled equations (2.1)-(2.5). For this
purpose we have solved those equations numerically for each point inside the
cloud, supposed to have a slab geometry and subjected to an external fixed
pressure $P$. Figs. (3)-(5) show some of the results obtained for different
values of $P$. The radiation field, illuminating the cloud from
both sides, has been held fixed at this stage to the fiducial value given by
eq. (2.17), and this particular choice will be referred as the {\it standard}
field. The two panels of each figure show the density and temperature
profiles as a function of the ratio $z/\ell$, where $z$ is the depth inside
the cloud and $\ell$ is the cloud size. Note that different curves refer to
different values of the HI column density and, therefore, of $\ell$; obviously,
solutions are symmetric about the point $z/\ell=0$ given that the symmetry
properties of the problem.
\par
As can be realized at a first glance to Figs. (3)-(5), the ionization structure
of the cloud strongly depends on the pressure of the external medium.
Following the discussion in \S~2.2, we have selected three different values
of the pressure $P$ that presumably may well represent the actual values,
namely $P=10^{-15},10^{-14},10^{-13}$~ergs~cm$^{-3}$.
We have disregarded the value $P=10^{-12}$~
erg/cm$^3$ because such a high pressure arises only in the static hot
halo model and in regions close to the disk, where the contamination
due to the galactic radiation field and local gaseous phenomena as
outflows can become dominant.
For $P=10^{-13}$~ergs~cm$^{-3}$, the cloud is initially
almost optically thin ($\ell=0.6$~pc, $N_H=2\times 10^{16}$~cm$^{-2}$) with
an ionization fraction $x\sim 0.42$ and a temperature $T\sim 9000$~K. When
the size is increased, self-shielding effects become more
and more
important, especially in the central parts of the cloud, and the ionization
and the temperature decrease steadily due to the decreased number of ionizing
photons available. The temperature structure will develop a
maximum not located at the cloud edge, but slightly displaced in the interior;
this is due to the fact that close to the edge, the temperature is determined
by the larger availability of electrons for cooling rather than by a decrease
in the number of ionizing photons. Eventually (last two curves), the
temperature
starts to fall down in an abrupt manner, while the ionization preserves its
regular pattern. Continuing to increase $N_H$, the cloud develops a cold
core, due to photon exhaustion in the partially ionized zone. This fact
confirms our simple
analytical estimates of \S~2, shown in Fig. 1. The characteristics of
this central dense ($n\sim 40 $~cm$^{-3}$), cold ($T\sim 20$~K) core are not
directly relevant to  the aim of this paper (anyway, some discussion is given
below).
The important point is, however, the {\it existence} of a critical
column density $N_H^c$, corresponding approximately to the curves with
the lowest
ionization fraction shown in each figure, after which a central condensation
completely shielded from the external radiation field is formed; in this case
$N_H^c\sim 7\times 10^{17}$~cm$^{-2}$. We will exploit this fact, as
explained below, to determine the distance to the HVCs.
\par
The same kind of behavior shown by the solutions for
$P=10^{-13}$~ergs~cm$^{-3}$, can be qualitatively recognized for the
two other pressure values adopted.
For $P=10^{-14}$~ergs~cm$^{-3}$ the ionization fraction is higher,
particularly when the medium
is still optically thin ($x\sim 0.8$), and  $N_H^c\sim 10^{19}$~cm$^{-2}$.
Finally, when the pressure is set to $P=10^{-15}$~ergs~cm$^{-3}$, the medium
is almost completely ionized and in order to  reach the  critical column
density the cloud size must be implausibly large ($\ell \simgt 30$~kpc)
if compared with any realistic model of the HVCs.
\par
The previous results can be summarized in a $N_H-\ell$ plane as the
one presented in Fig. (6). Each series of points represents the
$N_H-\ell$ relation for the curves previously discussed. According to
the previous discussion, the plane can be divided in two subregions
(Fig. 6): a lower one, in which  the ionization front is ``matter bounded''
and clouds do not have any cold core, and an upper one in which the front
is ``radiation bounded'' and a central cold core is found. Clouds located
in the upper part of the plane have thus a composite structure made up by a
neutral core surrounded by a warm envelope which constitutes an interface
with  the external, presumably hot,  medium. A simple approximate expression
for the ``critical'' curve $N_H^c(\ell)$ is found to satisfy
$$N_H^c(\ell)=N_1\sqrt{\ell_{pc}}~{\rm cm}^{-2},\eqno(3.1)$$
with $N_1= 3\times 10^{17}$~cm$^{-2}$. An analogous relation, which is
a very good approximation in the range $5\times 10^{15}\simlt P\simlt
10^{-13}$~ergs~cm$^{-3}$, between $N_H^c$ and $P$ is
$$N_H^c(\ell)=N_2 P_{-13}^{-1.1} {\rm cm}^{-2},\eqno(3.2)$$
where $N_2=7.4\times 10^{17}$ and $P_{-13}=P/10^{-13}$.
In addition to the case in which only the EBR field is considered,
the analougous results for the sum of the EBR + free-free ionizing
spectrum is also reported in Fig. 6. This addition is not modifying
the results in a substantial way because of the relatively low EM of the
hot gas, and hence of the low intensity of its radiation field.
Note that there are no open triangles in the case $P=10^{-15}$~ergs~cm$^{-3}$
because they are physically meaningless, indicating cloud sizes of the order of
$10^2$~kpc, larger than the halo itself; one may wonder about the
extragalactic nature of HVCs (which is excluded by almost any sort of data),
but the hypothesis of this paper should then be radically modified.
\par
It is natural to identify the core+interface structure found
with the observational evidences reported by many authors and discussed
in the Introduction, even if not much significance should be attached
to the precise values of the line widths since non thermal mechanisms may
concur to the line broadening; the important point is the
observed values, typically $\simgt 20$~km~s$^{-1}$, are higher than the ones
corresponding to our solutions.
At this point the suggested strategy  to determine the distance to the
HVCs becomes almost self-explanatory. In order to be as clear as possible
we put it in a schematic form:
\item{1.-} Using a cloud with a core+interface structure, measure its
$N_H$ in the interface; the measured $N_H$ will hence correspond to the
critical column density $N_H^c$;
\item{2.-} The value of $N_H^c$ must be located on the critical curve
(3.1). This gives the {\it linear} size of the interface, which, given
the existence of the core, must be exactly the $\ell$ corresponding to that
critical column density. By the way, pressure can be therefore immediately
determined through relation (3.2);
\item{3.-} Once the linear size of the interface $\ell$ is known, from
its angular diameter $\theta$, the distance $\Delta$ can be found through
the relation $\Delta=\ell/2\theta$.
\par
The procedure is quite straightforward if the cloud is observed to
have the required core+interface structure and column density measures
are available. Note that the actual characteristics and column density
of the core are not relevant to the distance determination. The mere
existence of the core bounds the column density of the partially ionized zone
(interface) to the appropriate critical density for the given pressure.
Some inferences on the physical state of the cores can nevertheless be made.
The very low temperature characterizing the cores in our model is to a
large extent determined by the only heating mechanism included apart from
the radiative one, \ie C ionization. Other heating mechanisms can of course
be foreseen; however, the non-detection of any HVCs in the IRAS maps (Wakker
\& Boulanger 1986), if interpreted as a low temperature of the dust, favours
the idea that HVCs are a cool environment.
However, the non-detection can be alternatively explained as due
to a low dust abundance, since dust can be heavily sputtered during
the expulsion process from the disk (Ferrara \etal 1991).
\par
Adding another step to our
procedure, we can also infer an approximate temperature of the core:
\item{4.-} Using the measured core $N_H$, from the determination of its
linear size $\ell_c$ which is possible knowing $\Delta$, we obtain an
average value of the density in the core $\langle n_c \rangle$, from
which $T_c$ can be derived from the equation of state.
\par
{}From a theoretical point of view, there are two main mechanisms which
can introduce some error in the
distance determined according to the scheme presented. The first one is
thermal conduction (Cowie \& McKee 1977; Draine \& Giuliani 1984; Begelman
\& Mc Kee 1990; McKee \& Begelman 1990; Ferrara \& Shchekinov 1993) whose
main effect is to broaden the
ionized zone. However, if clouds are formed from a thermal instability
in the fountain, by definition their size must be larger that the critical
wavelength at which thermal instabilities become stabilized by conduction
(Field 1965). Ferrara \& Shchekinov (1993) have demonstrated that the
dynamical effects of thermal conduction in such conditions are negligble.
Also conductive interfaces at the steady state tend to have a remarkably
flat temperature profile: for a sperical cloud, $T(r)=T_h(1-\ell/r)^{2/5}$,
where $r$ is the distance from the cloud center and $r\ge \ell$, and thus
most of the interface is at a temperature close to the one of the hot medium
$T_h$. In these condition virtually all the hydrogen is completely ionized
and the contribution of this gas to $N_H$  is negligible.
A second mechanism providing ionization is mechanical input by shocks.
The sound speed in the hot gas is
$$c_s=\left(\gamma P\over \mu m_p n_h\right)^{1/2}=126 \left({P_{-13}\over
n_{-3}}\right)^{1/2}~{\rm km~s}^{-1},\eqno(3.3)$$
where it has been assumed that the mean molecular weight $\mu=0.65$, $m_p$
is the proton mass, and
$n_{-3}=n_h/10^{-3}$ is the hot gas density; thus, by definition, some
HVCs move subsonically for the reference values of eq. (3.3).
The emission measure integrated on a path of 10 kpc for such a hot gas is
$E_m(n_{-3}=1)=10^{-2}$~cm$^{-6}$~pc, consistent with the one derived
by the ROSAT soft x-ray shadowing experiments (Burrows \& Mendenhall 1991;
Snowden \etal 1991); also the temperature derived by a fit to the data is
consistent with the estimate (3.3). Therefore, it appears that a relevant
fraction of the HVCs should not be affected by a shock, unless a large
transversal velocity component is present. However, shocks may have some
importance in the thermal balance of the core, and eventually may lead to
episodes of star formation, according to the calculations made by Dyson
\& Hartquist (1983). Viscous heating, ${\cal H}\propto \rho_h
v^3$ is not going to be important given the low density of the hot medium.
We conclude that the error on the distance determination due to
the effect of thermal conduction is negligible; in addition, for those
HVCs with velocity lower than $c_s$ specified by eq. (3.3), shocks are
not expected to appear.
\par
Probably more important are the possible errors related to real data.
It could be difficult to locate observationally the
position of the inner and outer edges of the interface in an accurate
manner. The inner edge is defined as the limit of the region with narrow
line profiles; however, this edge is often not regular. For the outer
edge, instead, it is crucial to resort to  high sensitivity data to detect
the low HI column density associated with the external regions of the
interface.
In addition, our assumption of a slab geometry can introduce some
additional error, particularly when the interface has dimensions much
larger than the core. This could be avoided by appropriately selecting the
cloud sample. With respect to the last point, we mention that in the
framework of the fountain model, HVCs are predicted to have a
a sheet-like shape (Kahn 1991) which would make the slab geometry
very appropriate.
\par
{}From the previous discussion it appears that the direct application
of the proposed method requires particular attention to both sample selection
and the estimation of possible errors. We plan to present these
results in a forthcoming publication; we stress, however, that a dedicated
observational study could serve this purpose much better.

\bigskip
\centerline{\sl 4. FURTHER IMPLICATIONS}
\smallskip
In the previous section we have presented a simple method based on the
photoionization structure of the HVCs which can be used in order to obtain
reliable distances to the clouds possesing a two-phase structure composed
by a warm ionized interface and a cold neutral core. However, our treatment
provides an opportunity to have some insight on some global properties of the
Galactic halo and the EBR. This is posssible if we compare the H$\alpha$
emission as predicted by our models with the one detected by
Kutyrev \& Reynolds (1989) (hereafter KR) and Songaila, Bryant \& Cowie (1989)
(hereafter SBC). The first authors
report the detection of H$\alpha$ emission from a HVC situated in Cetus
with a surface brightness $I_\alpha = 8.1 \pm 1.9\times10^{-2}$~R (where
one rayleigh corresponds to $10^6/4\pi$ photons cm$^{-2}$ s$^{-1}$ sr$^{-1}$);
Songaila \etal deduce  instead   $I_\alpha = 3\times10^{-2}$~R from two points
in a HVC belonging to the Complex C. These are the only detections we are
aware of and therefore, given the restricted sample, no definite conclusions
can be drawn; the comparison could be anyway of some interest.
\par
Assuming Case B recombination, the H$\alpha$ intensity
integrated on our model  slab cloud can be written as
$$I_\alpha=K \int_0^\ell n^2 x (x +h_c + x h_{He})T^{-0.96} dz~~{\rm
R},\eqno(4.1)$$
where $K=8.17\times 10^{-16}$, $n$ is the total gas density, $x$ is the
fractional ionization and $h_c, h_{He}$ are the abundances of C and He
relative to H, respectively. The results obtained applying eq. (4.1) to
the three models for low, intermediate and high pressure cases discussed in the
previous section are shown in Fig. 7 along with the oservational results.
It is immediately realized that the observed H$\alpha$ intensity is larger
than the theoretical one for all the pressure cases considered and the
discrepancy becomes larger as the pressure of the external medium is
increased.  The explanations for this dicotomy can all be found in two
basic possibilities, namely i) the intensity of the radiation field must
be larger than the one assumed, and/or ii) some mechanical energy input
must take place. As for the first point, it is difficult to conceive a
dramatically different value from the EBR intensity
given by eq. (2.17): in fact, it has been proven able to explain several
effects at high redshift and confirmed by observations in its high energy part.
If we exclude that any ionizing photon from the galaxy can escape the thick
Reynolds layer, the only remaining source of ionizing photons may come from
free-free emission of the hot gas confining the cloud. In the previous Section
we have seen that some modification to the ionization structure of the
cloud and to the interface size can be introduced by the consideration
of this contribution. However, the H$\alpha$ emission calculated
from the models including the radiation field from the local hot gas
show an enhancement only of a few percent with respect to the pure EBR field
case and certainly not sufficient to explain the observed excess.
\par
Thus we are left with the second possibility, \ie H$\alpha$ emission is
provided by recombinations following a shock wave created by the HVC/hot gas
interaction. Support for this hypothesis comes from the two following
evidences. First, the two clouds have velocities well in excess of the sound
speed in the hot medium (eq. [3.3]): for the KR cloud $\vert v_{LRS}\vert\sim
300$~km~s$^{-1}$, while for the SBC one $\vert v_{LRS}\vert\sim
140$~km~s$^{-1}$; second, the H$\alpha$ emission excess is correlated  with
the velocity as one would expect if part of the ionization is produced by
a shock. We conclude that the suggested method is not applicable to those
clouds in a straightforward manner since we have assumed that the ionization
of the clouds is due to the radiation field (EBR+free-free) only; at any rate
the distance method in such cases will provide at least a lower limit to
the distance. This could be particularly useful if absorption features towards
a background star are detected from the cloud. Finally, the calculated
H$\alpha$ emission (Fig. 7) provides a powerful test of the conditions in
which the proposed distance method can give reliable results, and clearly
indicates the possible presence of a shock.
\par
\vfill\eject
\vskip 3truecm
\centerline{\sl ACKNOWLEDGMENTS}
We thank S. M. Fall and an anonymous referee for useful comments.
\smallskip
\vfill\eject

\parindent=0pc
\parskip=6pt
\centerline{\sl REFERENCES}
\vskip 2pc

\ref Albert, C. E., Blades, J. C., Morton, D. C., Proulx, M. \& Lockman, F. J.
        1989, Structure and Dynamics of the Interstellar Medium, eds. G.
        Tenorio-Tagle, M. Moles \& J. Melnick (Berlin:Springer), 442

\ref Binney, J.  \& Tremaine, S. 1987, Galactic Dynamics,
     (Princeton:Univ.Press)

\ref Black J. H. 1981, MNRAS, 197, 555

\ref Blades, J. C., Wheatley, J. M., Panagia, N., Grewing, M., Pettini, M.
     \& Wamsteker, W. 1988, ApJ, 332, L75

\ref Bloemen, J. B. G. M. 1987, ApJ, 322, 694

\ref Boulares, A. \& Cox, D. P. 1990, ApJ, 365, 544

\ref Bowen, D. V. \& Blades, J. C. 1993, preprint

\ref Bregman, J. N. 1980, ApJ, 236, 577

\ref Bregman, J. N. \& Harrington, 1986, ApJ, 309, 833

\ref Brown, R. L. 1971, ApJ, 164, 387

\ref Brown, R. L. \& Gould, R. J. 1970, Phys. Rev. D, 1, 2252

\ref Burrows, D. N. \& Mendenhall, J. A. 1991, Nature, 351, 629

\ref Colgan, S. W. J., Salpeter, E. E. \& Terzian, Y.  1990, ApJ, 351, 503

\ref Cox, D. P. 1990, The Interstellar Medium in Galaxies, eds. H. A. Thoronson
     \& J. M. Shull, (Dordrecht:Kluwer), 181

\ref Cox, D. P. 1993, ESO/EIPC Workshop on Starburst Galaxies and Their
     Interstellar Medium, eds. J. Franco \& F. Ferrini, (Cambridge:Univ.Press),
     in press

\ref Cowie, L. L. \& McKee, C. F. 1976, ApJ, 209, L105

\ref Cowie, L. L. \& Songaila, A. 1986, ARA\&A, 24, 499

\ref Cowie L.L \& McKee, C. F. 1977, ApJ, 211, 135

\ref Cram, T. R. \& Giovanelli, R. 1976, A\&A, 48, 39

\ref Dalgarno, A. \& McCray, R. 1972, ARA\&A, 10, 375

\ref Danly, L., Albert, C. E. \& Kuntz, K. 1993, 3rd Annual Maryland Meeting,
     Back to The Galaxy, ed. NASA, in press

\ref de Boer, H. 1991, in The Interstellar Disk-Halo Connection in Galaxies,
     ed. H. Bloemen, (Dordrecht:Kluwer), 333

\ref de Boer, K. S.  \& Savage, B. D. 1984, ApJ, 136, L7

\ref Dickey, J. M. \& Lockman, F. J. 1990, ARA\&A, 28, 215

\ref Draine, B. T. \& Giuliani, J. L. 1984, ApJ, 281, 690

\ref Dyson, J. E. \& Hartquist, T. W. 1983, MNRAS, 203, 1233

\ref Ferrara, A. \& Ferrini, F., Franco, J. \& Barsella, B. 1991, ApJ, 381, 137

\ref Ferrara, A. \& Einaudi, G. 1992, ApJ, 395, 475

\ref Ferrara, A. 1993, ApJ, in press

\ref Ferrara, A. \& Shchekinov, Yu. 1993, ApJ, Nov. 20

\ref Field, G. B. 1965, ApJ, 142, 531

\ref Ferriere, K., Mac Low, M. M. \& Zweibel, E. G. 1991, 375, 239

\ref Herbstmeier, U., Kerp, J. Snowden, S. L. \& Mebold, U. 1993,
     Star Forming Galaxies and Their Interstellar Medium, eds. J. Franco
     \& F. Ferrini, (Cambridge:Univ. Press)

\ref Fransson, C. \& Chevalier, R. A. 5, ApJ, 296, 35

\ref Hartquist, T. W. \& Morfill, G. E. 1986, ApJ, 311, 518

\ref Houck, J. C. \& Bregman, J. N. 1990, ApJ, 352, 506

\ref Jahoda, K., McCammon, D., Dickey, J. M. \& Lockman F. J. 1985, ApJ, 290,
229

\ref Kahn, F. D. 1991, The Interstellar Disk-Halo Connection
     in Galaxies, ed. H. Bloemen, (Dordrecht:Kluwer), 1

\ref Kulkarni, V. P. \& Fall, S. M. 1993, ApJ, in press (Aug. 20 issue)

\ref Kutyrev, A. S. \& Reynolds, R. J. 1989, ApJ, 344, L9

\ref Landini, M \& Monsignori Fossi, B. C. 1990, AASS, 82, 229

\ref Li, F. \& Ikeuchi, S. 1992, 390, 405

\ref Mac Low, M. M. \& Mc Cray, R. 1988, ApJ, 324, 776

\ref Mac Low, M. M. , Mc Cray, R. \& Norman, M. L. 1989, ApJ, 337, 141

\ref McKee, C. F. \& Begelman, M. C. 1990, ApJ, 358, 392

\ref Madau, P.  1992, ApJ, 389, L1

\ref Mc Cammon, D. \& Sanders, W. T. 1990, ARA\&A, 28, 657

\ref Mirabel, I. F. 1989, Structure and Dynamics of the Interstellar Medium,
     eds. G.  Tenorio-Tagle, M. Moles \& J. Melnick (Berlin:Springer), 396

\ref Morrison, R. \& Mc Cammon, D. 1983, ApJ, 270, 119

\ref Nordgren, T. E., Cordes, J. M. \& Terzian, Y. 1993, preprint

\ref Norman, M. L, 1993, 3rd Annual Maryland Meeting, Back to The Galaxy, ed.
     NASA, in press

\ref Norman, C. A. \& Ikeuchi, S. 1989, ApJ, 345, 372

\ref Norman, C. A. \& Panagia, N. 1991, The Interstellar Disk-Halo Connection
     in Galaxies, ed. H. Bloemen, (Dordrecht:Kluwer), 325

\ref Reynolds, R. J. 1990, ApJ, 349, L17

\ref Reynolds, R. J. 1993, 3rd Annual Maryland Meeting, Back to The Galaxy,
     ed.  NASA, in press

\ref Rohlfs, R., Herbstmeier, U., Mebold, U., \& Winnberg, A. 1989, A\&A, 211,
402

\ref Sargent, W. L. W., Young, P. J., Boksenberg, A., Carswell, R. F. \&
     Whelan, J. A. J. 1979, ApJ, 230, 49

\ref Schwartz, D. A. 1979, X-Ray Astronomy, eds. W. A. Baity \& L. E. Peterson,
     (Oxford:Pergamon), 453

\ref Sembach, K. R., Savage B. D., \& Massa, D. 1991, ApJ, 372, 81

\ref Shapiro, P. R. \& Benjiamin, R. A. 1993,
     Star Forming Galaxies and Their Interstellar Medium, eds. J. Franco
     \& F. Ferrini, (Cambridge:Univ.Press)

\ref Shapiro, P. R. \& Field, G. B. 1976, ApJ, 205, 762

\ref Shull, J.M. \& Van Steenberg, M. E. 1985, ApJ, 298, 268

\ref Snowden, S. L., Mebold, U., Hirth, W., Herbstmeier, U. \& H. M. M.
     Schmitt 1991, Science, 292, 1529

\ref Songaila, A., Cowie, L. L.  \& Weaver, H. 1988, ApJ, 329, 580

\ref Songaila, A., Bryant, W. \& Cowie, L. L. 1989, ApJ, 345, L71

\ref Spitzer, L. 1956, ApJ, 124, 20

\ref Spitzer, L. 1978, Physical Processes in the Interstellar Medium,
     (New York:Wiley)

\ref Spitzer, L. 1990, ARA\&A, 28, 71

\ref Tenorio-Tagle, G., Ro\.zyczka, M. \& Bodenheimer, P. 1990, A\&A, 237, 207

\ref Terasawa, N. 1992, ApJ, 392, L15

\ref Tomisaka, K., 1990, ApJ, 361, L5

\ref van Woerden 1993, Luminous High-Latitude Stars, ed. D. D. Sasselov,
     PASP Conference Series, in press

\ref Viala, Y. P. \& Horedt, Gp. 1974, A\&A, 33, 195

\ref Wakker, B. P.  1989, Thesis, Groningen University

\ref Wakker, B. P. \& Boulanger, F. 1986, A\&A, 170, 84

\ref Wakker, B. P. \& Schwarz, U. J. 1991, A\&A, 250, 484

\ref Wakker, B. P. 1991, The Interstellar Disk-Halo Connection
     in Galaxies, ed. H. Bloemen, (Dordrecht:Kluwer), 27

\vfill\eject

\bigskip
\centerline{\sl FIGURE CAPTIONS}
\smallskip
\vskip0.2in
{\bf Figure 1} Ionization structure of the cloud partially ionized zone.
The curves show the solutions of eq. (2.15); the numbers refer to different
values of the ionization parameter  $\Xi=P/J_0\zeta$; $z$ is measured from
the cloud edge.
\vskip0.1in
{\bf Figure 2} Pressure isocontours for a static, isothermal, hot halo;
the temperature and pressure at the disk are $T=10^6$~K, $P_\odot=3.5\times
10^{-12}$~erg~cm$^{-3}$.
\vskip0.1in
{\bf Figure 3} Ionization and temperature structure for a cloud of size $\ell$
exposed to the {\it standard} EBR field given by eq. (2.17); the pressure is
constant at a value  $P=10^{-13}$~erg~cm$^{-3}$. Each line refers to a
different value of the column density $N_H$ which is in the range $3\times
10^{16} \le N_H \le 7.5\times 10^{17}$~cm$^{-2}$.
\vskip0.1in
{\bf Figure 4} Same as Fig. 3, but for $P=10^{-14}$~erg~cm$^{-3}$, and
$5\times 10^{16} \le N_H \le 8.5\times 10^{18}$~cm$^{-2}$.
\vskip0.1in
{\bf Figure 5} Same as Fig. 3, but for $P=10^{-15}$~erg~cm$^{-3}$, and
$1\times 10^{15} \le N_H \le 1.4\times 10^{18}$~cm$^{-2}$.

\vskip0.1in
{\bf Figure 6} Hydrogen column density of the cloud as a function of its
size; the numbers refer to different values of the pressure in erg~cm$^{-3}$.
{\it Solid} triangles refer to the {\it standard} EBR field given by eq.
(2.17),
{\it open} triangles are for the composite field EBR+free-free from hot gas.
The upper dashed part describes the region of the parameter space in which
the clouds develop a core+interface structure in the EBR field.
\vskip0.1in
{\bf Figure 7} Calculated H$\alpha$ intensity, $I_\alpha$, from the cloud
interface as a function of cloud size for the three models of Fig. 6; the
numbers refer to different
values of the pressure in erg~cm$^{-3}$. Also shown (horizontal lines) are
the observed values for two HVCs as given by RK and SBC.
\vfill\eject
\bye